\documentclass[12pt]{article}

\usepackage{amsmath}
\usepackage{amssymb}
\usepackage{epsfig}
\usepackage{graphicx}
\usepackage{pstricks}
\usepackage{pst-coil}

\newcommand{\Sol}  {\textrm{sol}}

\newcommand{\Atm}  {\textrm{atm}}

\begin{document}

%\begin{titlepage}
\vspace*{-1cm}
\phantom{hep-ph/***}
\hfill{IFIC/09-03}
\vskip 2.5cm
\begin{center}
{\Large\bf Tri-Bimaximal lepton mixing with $A_4 \ltimes (Z_2)^3$\\
\vskip .3cm}
\end{center}
\vskip 0.5  cm
\begin{center}
{\large Stefano Morisi}~$^{a)}$\footnote{e-mail address: morisi@ific.uv.es}
\\
\vskip .2cm
$^{a)}$~AHEP Group, Institut de F\'{\i}sica Corpuscular --
  C.S.I.C./Universitat de Val{\`e}ncia \\
  Edificio Institutos de Paterna, Apt 22085, E--46071 Valencia, Spain
\vskip 0.7cm
\end{center}

\begin{abstract}
We consider a model %that yields to tri-bimaximal lepton mixing 
based on $A_4\ltimes (Z_e\times Z_\mu \times Z_\tau)$ 
flavor symmetry for Tri-Bimaximal lepton mixing with two scalar $A_4$-triplets whose  vacuum expectation values have the same alignment.
%We study the  alignment of the vacuum expectation values of the Higgs doublets.
Neutrino masses are generated through effective dimension five Weinberg operators.
Charged leptons mass hierarchies
are generated with an additional  Froggatt-Nielsen $U(1)_F$ flavor symmetry under which right-handed leptons are charged.

\end{abstract}
%\end{titlepage}
\setcounter{footnote}{0}


\vskip2truecm

\section{Introduction}
A successful phenomenological ansatz for leptons has been proposed by
Harrison, Perkins and Scott (HPS) and is given
by~\cite{Harrison:2002er}
\begin{equation}
\label{eq:HPS}
U_{\textrm{HPS}} = 
\left(\begin{array}{ccc}
\sqrt{2/3} & 1/\sqrt{3} & 0\\
-1/\sqrt{6} & 1/\sqrt{3} & -1/\sqrt{2}\\
-1/\sqrt{6} & 1/\sqrt{3} & 1/\sqrt{2}
\end{array}\right)
\end{equation}
which corresponds to 
$\tan^2\theta_{\Atm}=1$, $\sin^2\theta_{\textrm{Chooz}}=0$ and $\tan^2\theta_{\Sol}=0.5$, 
providing a good first approximation to the values indicated by
current neutrino oscillation data \cite{Schwetz:2008er,Fogli:2005cq}.

There are many studies to derive the so called Tri-Bimaximal (TB) lepton mixing in eq.\,(\ref{eq:HPS}) by means of non-abelian
groups like $A_4$\,\cite{tbm1,tbm2}, $T'$\,\cite{Feruglio:2007uu,Chen:2007afa}, $S_4$\,\cite{Lam:S4natural,Bazzocchi:2008ej,Ishimori:2008fi} and $\Delta(27)$\,\cite{deMedeirosVarzielas:2006fc}.
However all models with non-abelian flavor symmetry $G_f$ giving TB lepton mixing, need at least
two scalar fields whose vacuum expectation values (vevs) 
break $G_f$ into distinct subgroups of $G_f$. In particular $A_4$-based models  need two $A_4$-triplets, for instances
$\varphi$ and $\chi$, whose vevs are $\langle\varphi_1\rangle=\langle\varphi_2\rangle=\langle\varphi_3\rangle$ and 
$\langle\chi_1\rangle\ne\langle\chi_2\rangle=\langle\chi_3\rangle=0$. We say that  
 $\varphi$ and $\chi$ have distinct vevs {\it alignments}. When $\varphi$ and $\chi$  take vevs $A_4$ breaks respectively into $Z_2$ and $Z_3$.
Different papers \cite{Ma:2008ym,Grimus:2008tt,Zee:2005ut,Ma:2004zv,Hirsch:2008mg} have recently emphasized 
the difficulty to have distinct vevs alignments\footnote{Also  called the {``misalignment problem''}\cite{Feruglio:2007uu}.}.  
In order to account for such alignments, extra-dimensions~\cite{Altarelli:2005yp,Kobayashi:2008ih},
 supersymmetry~\cite{Babu:2005se,Altarelli:2005yx} or Wilson lines   \cite{Seidl:2008yf}
 have been invoked.
Recently has been studied models for TB mixing without vevs alignments \cite{Ma:2008ym,Hirsch:2008mg,Grimus:2008vg}  
where  the flavor symmetry is softly broken.

In this paper we study the possibility to generate TB mixing in the lepton sector with two $A_4$ scalar triplets. 
Their vevs break spontaneously $A_4$ into $Z_2$ and have the same alignment. 
We do not have $A_4$ scalar triplets whose vevs are misaligned to get TB mixing.
We need to enlarge the flavor group $A_4$ to the group defined as $G_f=A_4\ltimes (Z_e\times Z_\mu \times Z_\tau)$,
that is the semidirect product \footnote{See for instance \cite{Seidl:2008yf} for a short introduction.} 
of $A_4$ with the group $(Z_2)^3\equiv(Z_e\times Z_\mu \times Z_\tau)$.
As in \cite{Zee:2005ut}, we would like to be as minimal as possible generating charged lepton masses by dimension four operators
and constructing the neutrino masses by dimension five Weinberg operators \cite{Weinberg:1979sa}, without enter into the details of the 
particular dynamical model that generates such a operators.

In the next section we introduce the $A_4$ group, in section\,3 we give the main feature of the model based on $G_f$ flavor symmetry, 
in section 4 we study the Higgs potential and then 
in section 5 we give the conclusions.

\section{The group $A_4$}
$A_4$ is 
the finite group of the even permutations of four objects (for a short
introduction to $A_4$, see for instance \cite{Altarelli:2005yx} and
references therein). All the 12 elements of the group can be generated from two elements $S$ and $T$ 
that satisfy the following defining rules
\begin{equation}
S^2=T^3=(ST)^3=\mathcal{I}.
\end{equation}
\begin{table}[t]
\begin{center}
\begin{tabular}{|l|c|c|c|c|}
\hline 
&$C_1=\{I\}$ & $C_2=\{T\}$ & $C_3=\{T^2\}$ & $C_4=\{S\}$ \\
\hline
1&1 & 1&1 &1\\
\hline
$1'$&1 &$\omega$ & $\omega^2$&1\\
\hline
$1''$&1 &$\omega^2$ & $\omega$&1\\
\hline
3&3 &0 & 0&$-1$\\
\hline
\end{tabular}\caption{Character table of $A_4$ where $C_i$ are the different classes and $\omega^3\equiv 1$. }\label{tab0}
\end{center}
\end{table}
Since $A_4$ group has four equivalence classes, 
there are four irreducible representations, one
triplet $3$ and three singlets $1,1',1''$, see table\,(\ref{tab0}).  
The product of the singlets is equal to the $Z_3$ product, then $1'\times 1''\sim 1$, $1'\times 1'\sim 1''$ and so on.
The product of two triplet $3\times 3$ contains the three singlets and two triplets. 
The form of the irreducible representations contained into the product of two triplets, depends from the 
choice of the structure of the generators $S$ and $T$. Here we take the basis where $T$ is diagonal
\begin{equation}\label{st}
T=
\left(\begin{array}{ccc}
 1  & 0   &0  \\
0   & \omega  & 0 \\
0   & 0  & \omega^2
\end{array}\right),\qquad
S=\frac{1}{3}
\left(\begin{array}{ccc}
 -1  & 2   &2  \\
 2  & -1  &2  \\
 2  & 2  & -1
\end{array}\right).
\end{equation}
In such a basis the product of two triplets $\chi$ and $\varphi$
is given by
\begin{equation}\label{prod}
\begin{array}{c}
1\sim(\chi\varphi)  =(\chi _1\varphi_1+\chi_2\varphi_3+\chi_3\varphi_2),\\
1'\sim(\chi\varphi)' =(\chi_3\varphi_3+\chi_1\varphi_2+\chi_2\varphi_1),\\
1''\sim(\chi\varphi)''=(\chi_2\varphi_2+\chi_1\varphi_3+\chi_3\varphi_1),\\
\\
3_s\sim
\left(
\begin{array}{c}
2\chi_1\varphi_1-\chi_2\varphi_3-\chi_3\varphi_2\\
2\chi_3\varphi_3-\chi_1\varphi_2-\chi_2\varphi_1\\
2\chi_2\varphi_2-\chi_1\varphi_3-\chi_3\varphi_1\\
\end{array}
\right),\,
3_a\sim
\left(
\begin{array}{c}
\chi_2\varphi_3-\chi_3\varphi_2\\
\chi_1\varphi_2-\chi_2\varphi_1\\
\chi_1\varphi_3-\chi_3\varphi_1
\end{array}
\right).
\end{array}
\end{equation}
We observe that in the basis of eq.\,(\ref{st}) the product $3 \times 3$ is different from $3 \times \bar{3}$ since $T$ is complex.
For instance, the product of $\chi$ and $\bar{\varphi}$ is given by eq.\,(\ref{prod}) with $\varphi_2\leftrightarrow \varphi_3$.

\section{The model}
Consider the model defined in Table\,(\ref{tab1}).
The left-handed doublets transform as a triplet of $A_4$, namely $L=(L_e,L_\mu,L_\tau)$ and the right-handed fields $e^c$, $\mu^c$, $\tau^c$ as different singlets
of $A_4$. Each right-handed field $l^c_a$ is charged under the corresponding $Z_{2a}$ with $a=e,\mu,\tau$. 
We have two  Higgs doublets $h$  and $\xi$ that are singlets of $A_4$ and they differ 
for an extra abelian symmetry $Z_3$. There is an $A_4$-triplet $\varphi=(\varphi_1,\varphi_2,\varphi_3)$ of Higgs doublets. 
We have also one more $A_4$-triplet of Higgs doublets $\phi=(\phi_1,\phi_2,\phi_3)$ where 
each %component of the triplet $\phi$, namely 
scalar field $\phi_a$ transforms 
with respect to $Z_{2a}$. %This is possible since $\phi$ belong to a group $G_f$ that is bigger than $A_4$.
As in \cite{Grimus:2005rf,Mohapatra:2006pu} the $(Z_2)^3$ symmetries glue each $l_i^c$ with the corresponding $\phi_i$. 
As we will show below, $(Z_2)^3$  remove off-diagonal terms in the charged lepton sector,
and as a consequence the charged lepton mass matrix is diagonal. 

\begin{table}[t]
\begin{center}
\begin{tabular}{|c|c|c|c|c||c|c|c|c|c|c|}
\hline
&$L_{}$&$e^c$&$\mu^c$&$\tau^c$   &$\xi$&$h$ &$\varphi$& $\phi_1$ & $\phi_2$ & $\phi_3$\\
\hline
%$SU(2)$&$2$&$1$&$1$&$1$  &$2$&$2$&$2$\\
$A_4$&$3$&$1$&$1'$&$1''$&$1$&1&3&\multicolumn{3}{c|}{3}\\\hline
$Z_{2e}$   &$+$&$-$&$+$&$+$&$+$&$+$&$+$&$-$&$+$&$+$\\\hline
$Z_{2\mu}$ &$+$&$+$&$-$&$+$&$+$&$+$&$+$&$+$&$-$&$+$\\\hline
$Z_{2\tau}$&$+$&$+$&$+$&$-$&$+$&$+$&$+$&$+$&$+$&$-$\\\hline
$Z_3$& $\omega$  & $\omega^2$  &  $\omega^2$    & $\omega^2$ &1     &$\omega$ &1&1&1&1\\
\hline
\end{tabular}\caption{Lepton and scalar multiplet structure of our 
model, see text. }\label{tab1}
\end{center}
\end{table}

The $G_f\times Z_3$ invariant Lagrangian reads
\begin{eqnarray}
\label{eq:LL}
\mathcal{L}&=&{y_{e}}L_e e^c\phi_1+y_\mu L_\mu  \mu^c\phi_2+y_\tau L_\tau  \tau^c\phi_3+
%\nonumber\\ &&+
\frac{y_a}{\Lambda} (L L )\xi h +\frac{y_b}{\Lambda} (L L \varphi) h,
\end{eqnarray}
where $(LL)$ means the $A_4$-singlet contained into $3\times 3$, $(LL\phi)$ the $A_4$-singlet contained into $3\times 3\times 3$ and $\Lambda$
is the cut-off scale.
According to the $A_4$ symmetry also the following terms are allowed (see eq.\,(\ref{prod}))
\begin{eqnarray}
&&y_{e}  (L_\mu  e^c\phi_3 + L_\tau  e^c\phi_2 )+
y_\mu  (L_e \mu^c\phi_3+L_\tau H \mu^c\phi_1)+
y_\tau (L_e H \tau^c\phi_2+L_\tau H \tau^c\phi_1),
%
%&&y_{e}  (L_\mu  e^c\phi_3 + L_\tau  e^c\phi_2 )+\nonumber\\
%&&y_\mu  (L_e \mu^c\phi_3+L_\tau H \mu^c\phi_1)+\nonumber\\
%&&y_\tau (L_e H \mu^c\phi_2+L_\mu H \mu^c\phi_1)
\end{eqnarray}
but these terms are not invariant under the $Z_e\times Z_\mu \times Z_\tau$ symmetry. 

When the $\phi$'s Higgs doublets take vevs as below
\begin{equation}\label{vev1}
\langle \phi_1\rangle = \langle \phi_2\rangle = \langle \phi_3\rangle=v,
\end{equation}
the charged lepton mass matrix is diagonal with masses proportional to the yukawa couplings $y_e, \, y_\mu$
and $y_\tau$. 
We can generate the charged lepton mass hierarchies  assuming that the right-handed charged fields transform with respect to an extra Froggatt-Nielsen (FN)
symmetry $U(1)_F$.
Assuming for instance the FN charges $F$ to be $F(e^c)=4$,  $F(\mu^c)=2$, $F(\tau^c)=0$ and   
a flavon field $\theta$ with $F(\theta)=-1$, then $y_e\sim \lambda^4$, $y_\mu\sim \lambda^2$ and $y_\tau\sim 1$ where $\lambda\sim\langle \theta \rangle/\Lambda$. 
We can reproduce the correct lepton mass hierarchies if 
$\lambda\approx 0.22$.

%Our approach is completely different from the model of ref.\,\cite{Grimus:2008vg}  where the fermion mass hierarchies is related to the 
%different vevs of $\phi_1$, $\phi_2$ and $\phi_3$. In \cite{Grimus:2008vg} as well as in \cite{Grimus:2008tt} has been showed that 
%$\langle \phi_1\rangle \ll \langle \phi_2\rangle \ll \langle \phi_3\rangle$ is possible by assuming {\it soft-breaking} terms 
%of dimension two in the scalar potential like $\phi_i^\dagger \phi_i$ and/or $\phi_i^\dagger \phi_j$.

When the scalar Higgs fields $\xi$ and $h$ take vevs $ \langle \xi\rangle=u$ 
and $ \langle h\rangle=t$ and assuming  
\begin{equation}\label{vev2}
\langle \varphi_1\rangle = \langle \varphi_2\rangle = \langle \varphi_3\rangle=k,
\end{equation}
the Majorana neutrino mass matrix takes the form
\begin{equation}
M_\nu=\left(\begin{array}{ccc}
a+2b & -b& -b\\
-b &2b& a-b\\
-b & a-b & 2b
\end{array}\right),
\end{equation}
where $a=y_a u t/\Lambda$ and $b=y_b k t/\Lambda$. 
We will show in the next section that the vevs alignments of eqs.\,(\ref{vev1}) and (\ref{vev2}) are natural in our model. 
The matrix $M_\nu$ is $\mu\leftrightarrow\tau$ invariant and $M_{\nu_{11}}+M_{\nu_{13}} =  M_{\nu_{22}} +M_{\nu_{23}}$.
Therefore the atmospheric angle is maximal, the reactor angle is zero and the solar angle trimaximal, see for instance
 \cite{Hirsch:2008mg}. We have three different  eigenvalues $m_1=a+3b$,
$m_2=a$ and $m_3=-a+3b$ and it is possible to reproduce  the ratio $r=\Delta m_{\textrm{sol}}^2/\Delta m_{\textrm{atm}}^2$, see for instance \cite{Altarelli:2005yp}.

\section{The scalar potential}
In our model there are 8 Higgs doublets that belong to different representations of $G_f$. 
We have assumed the vev of the two $A_4$-triplets to be  equally aligned, that is  $\langle \phi_1\rangle=\langle \phi_2\rangle=\langle \phi_3\rangle$
and $\langle \varphi_1\rangle=\langle \varphi_2\rangle=\langle \varphi_3\rangle$.
We have showed that these alignments together with the $(Z_2)^3$ symmetry give  TB mixing. 
In \cite{Altarelli:2005yp} has already been showed that such a alignment is a solution of the most generic
$A_4$ invariant potential.
Here we study the Higgs potential and we show that the above 
alignment is one possible minimum.

The scalar potential invariant under  $G_f\times Z_3$ is  
\begin{eqnarray}
V&= &\mu_1 h^\dagger h+\mu_2\xi^\dagger \xi+ \mu_3(\phi^\dagger \phi)+ \mu_4(\varphi^\dagger \varphi)+  \nonumber\\
&+& \sum_{a,b=r}\lambda_{ab}(\phi^\dagger \phi)_a(\phi^\dagger \phi)_b +
\sum_{a,b=r}\lambda'_{ab}(\varphi^\dagger \varphi)_a(\varphi^\dagger \varphi)_b +
\sum_{a,b=r}\lambda''_{ab}(\phi^\dagger \phi)_a(\varphi^\dagger \varphi)_b +\nonumber\\
&+&\lambda_1 (\phi^\dagger \phi)\xi^\dagger \xi+\lambda_2(\phi^\dagger \phi)h^\dagger h+
\lambda_3\xi^\dagger \xi\xi^\dagger \xi + \lambda_4h^\dagger h h^\dagger h+ \lambda_5\xi^\dagger \xi h^\dagger h+\nonumber\\
&+&\lambda_6 (\varphi^\dagger \varphi)\xi^\dagger \xi+\lambda_7 (\varphi^\dagger \varphi)h^\dagger h+
\mbox{h.c.}
\end{eqnarray}
where $r=1,1',1'',3_s,3_a$. We assume the vevs of the Higgs doublets to point in the same direction of $SU(2)$ and, for simplicity all the vevs to be real.
We define 
$$\langle \varphi_i\rangle=k_i,\quad \langle \phi_i\rangle=v_i,\quad \langle \xi \rangle=u ,\quad \langle h \rangle=t ,  $$
then the scalar potential reads
\begin{eqnarray}
&V_{min}&=\mu_1 t^2+\mu_2 u^2+\mu_3 (v_1^2 +v_2^2 +v_3^2 )+ \mu_4 (k_1^2 +k_2^2 +k_3^2 )+  \nonumber\\
&+& \lambda_{11}(v_1^2 +v_2^2 +v_3^2 )^2+\lambda_{1'1''}(v_3v_2 +v_1v_2 +v_3v_1 )^2+    \nonumber\\ 
&+&\lambda_{3_s3_s}\left[(2v_1^2 -v_2^2 -v_3^2 )^2+2(2v_3 v_2 -v_1 v_3 -v_2 v_1 )^2 \right]+\nonumber\\
&+&\lambda_{3_a3_a}\left[(v_3^2 -v_2^2 )^2- 2(v_1v_3-v_1v_2 )^2\right]+\nonumber\\
&+& \lambda'_{11}(k_1^2 +k_2^2 +k_3^2 )^2+\lambda'_{1'1''}(k_3k_2 +k_1k_2 +k_3k_1 )^2+    \nonumber\\ 
&+&\lambda'_{3_s3_s}\left[(2k_1^2 -k_2^2 -k_3^2 )^2+2(2k_3 k_2 -k_1 k_3 -k_2 k_1 )^2 \right]+\nonumber\\
&+&\lambda'_{3_a3_a}\left[(k_3^2 -k_2^2 )^2- 2(k_1k_3-k_1k_2 )^2\right]+\nonumber\\
&+& \lambda''_{11}(v_1^2 +v_2^2 +v_3^2 ) (k_1^2 +k_2^2 +k_3^2 ) +    %\nonumber\\   &+&
\lambda''_{1'1''}(v_3v_2 +v_1v_2 +v_3v_1 )(k_3k_2 +k_1k_2 +k_3k_1 )+    \nonumber\\ 
&+&\lambda''_{3_s3_s}((2v_1^2 -v_2^2 -v_3^2 )(2k_1^2 -k_2^2 -k_3^2 )+\nonumber\\
&&\quad+2 (2v_3 v_2 -v_1 v_3 -v_2 v_1 ) (2k_3 k_2 -k_1 k_3 -k_2 k_1 ) )+\nonumber\\
&+&\lambda''_{3_a3_a}\left[(v_2^2 -v_3^2 )(k_2^2 -k_3^2 )- 2(v_1v_3-v_1v_2 )(k_1k_3-k_1k_2 )\right]+\nonumber\\
&+& \lambda_1 (v_1^2 +v_2^2 +v_3^2 ) u^2+
\lambda_2 (v_1^2 +v_2^2 +v_3^2 ) t^2+\lambda_3 u^4+\lambda_4 t^4+\lambda_5 u^2t^2+\nonumber\\
&+&\lambda_6 (k_1^2 +k_2^2 +k_3^2 ) u^2+\lambda_7 (k_1^2 +k_2^2 +k_3^2 ) t^2\,.
\end{eqnarray}
A possible solution of the system $\partial V/\partial x_i=0$ with $x_i=v_i,\,k_i,\,u,\,t$  is 
%By studing the system $\partial V/\partial x_i=0$ with $x_i=v_i,\,k_i,\,u,\,t$ we have that 
\begin{equation}\label{sol12}
v_1=v_2=v_3=v\ne 0,\qquad k_1=k_2=k_3 =k\ne 0,%\quad\textrm{with}
\qquad t\ne 0,\qquad u\ne 0, 
\end{equation}
%is a possible solution where for instance
with
\begin{eqnarray}
v&=&\{\lambda_5\lambda_6\lambda'' \mu_1-
2\lambda_3\lambda_7\lambda''\mu_1-
2\lambda_4\lambda_6\lambda''\mu_2+
\lambda_5\lambda_7\lambda'' \mu_2+
6\lambda_4\lambda_6^2 \mu_3
-6\lambda_5\lambda_6\lambda_7\mu_3+\nonumber\\
&+&6\lambda_3\lambda_7^2\mu_3-
24\lambda_3\lambda_4\lambda'\mu_3+
6\lambda_5^2\lambda'\mu_3+
4\lambda_3\lambda_4\lambda'' \mu_4-
\lambda_5^2\lambda'' \mu_4+\nonumber\\
&+&\lambda_2(-\lambda_6^2\mu_1+4\lambda_3\lambda' \mu_1+\lambda_6\lambda_7 \mu_2-2\lambda_5\lambda' \mu_2+\lambda_5\lambda_6 \mu_4-\lambda_3\lambda_7 \mu_4)+\nonumber\\
&+&3\lambda_1(\lambda_6\lambda_7\mu_1-2\lambda_5\lambda' \mu_1-\lambda_7^2 \mu_2+4\lambda_4\lambda' \mu_2-2\lambda_4\lambda_6 \mu_4+\lambda_5\lambda_7 \mu_4)\}
\,/ \nonumber\\
&&\{ 
9\lambda_1^2\lambda_7^2 
-36\lambda_1^2\lambda_4\lambda' 
+\lambda_2^2(\lambda_6^2  -4 \lambda_3\lambda') 
-4\lambda(\lambda_4 \lambda_6^2- \lambda_5\lambda_6\lambda_7+ \lambda_3\lambda_7^2-4 \lambda_3\lambda_4\lambda' +\lambda_5^2\lambda')+\nonumber\\ 
&+&12\lambda_1\lambda_4 \lambda_6\lambda'' 
-6\lambda_1\lambda_5 \lambda_7\lambda'' 
-4\lambda_3\lambda_4 {\lambda''}^2 
+\lambda_5^2 {\lambda''}^2 +\nonumber\\
&-&2\lambda_2(3\lambda_1 \lambda_6\lambda_7- 6\lambda_1\lambda_5\lambda'+ \lambda_5\lambda_6\lambda''-2 \lambda_3\lambda_7\lambda'') 
\},
\end{eqnarray}
with $\lambda=9(\lambda_{11}+\lambda_{1'1''})$, $\lambda'=9(\lambda'_{11}+\lambda'_{1'1''})$ and $\lambda''=9(\lambda''_{11}+\lambda''_{1'1''})$. 
Similar relations can be found for $k$, $u$ and $t$.
We have verified that there is a non-vanishing portion of the parameter space where the Hessian matrix $\partial^2 V/\partial x_i \partial x_j$
has positive eigenvalues for the solution in eq.\,(\ref{sol12}).

\section{Conclusions}
We have studied a model for TB lepton mixing. The model is based on the flavor symmetry  $A_4 \ltimes (Z_2)^3$. Most of the models
based on $A_4$ need two scalar flavor-triplets that take vevs with different alignments in order to give TB mixing. The reason is that one of the two 
scalar triplet breaks $A_4$ into $Z_2$ in the neutrino sector, while the second scalar triplet breaks  $A_4$ into $Z_3$ in the charged lepton sector.
The misalignment between the two sectors yields to the large TB mixing.
However it has been shown that vevs misalignment is not a natural solution of the scalar potential without introducing supersymmetry or extra-dimensions.

In our model the vevs of the two scalar flavor-triplets interacting respectively with the neutrino sector and the charged lepton sector,
have the same alignment. Unwanted off-diagonal terms in the charged lepton
mass matrix are removed by means of the $(Z_2)^3$ symmetry. Since the groups $A_4$ and  $(Z_2)^3$ do not commute, the flavor group is bigger than $A_4$ and is given 
by their semidirect product, namely  $A_4 \ltimes (Z_2)^3$. This method to remove off-diagonal terms in the charged sector has  already been used in 
\cite{Grimus:2008tt,Grimus:2008vg,Grimus:2005rf,Mohapatra:2006pu} where the basic groups are $S_3$ and $\Delta(27)$.  
In \cite{Grimus:2008tt,Grimus:2008vg,Grimus:2005rf,Mohapatra:2006pu}  
the fermion mass hierarchies are related to the 
different vevs of $\phi_1$, $\phi_2$ and $\phi_3$ and it has been showed that 
$\langle \phi_1\rangle \ll \langle \phi_2\rangle \ll \langle \phi_3\rangle$ is possible by assuming {\it soft-breaking} terms 
of dimension two in the scalar potential like $\phi_i^\dagger \phi_i$ and/or $\phi_i^\dagger \phi_j$.
Our approach is completely different because we have $A_4$ then 
the fermion masses are proportional to the yukawa couplings. 
%Since left-handed fields are assigned to a triplet $A_4$ representation and right-handed fields to $1$, $1'$ and $1''$.
We can generate fermion mass hierarchies by adding an extra FN symmetry and we do not require to softly break the flavor  symmetry.
We have studied the $G_f\times Z_3$ invariant Higgs potential. We have shown that the solution with all the $\phi$'s and $\varphi$'s acquiring equal vevs is a possible 
minimum. 
Neutrino get masses from a dimension-five Weinberg operators. We do not enter into the details of the dynamical mechanism that generate light neutrino masses. 
%While in  \cite{Grimus:2008tt} TB mixing is obtained by means of type-I seesaw mechanism where more than three
%right-handed neutrino are required. 

\section*{Acknowledgments}
Work supported by MEC grant FPA2008-00319/FPA, by European Commission Contracts
MRTN-CT-2004-503369 and ILIAS/N6 RII3-CT-2004-506222.

\end{document}